\documentclass[%
 aip,
cp,  
 amsmath,amssymb,
 reprint,%
]{revtex4-2}

\expandafter\let\csname equation*\endcsname\relax
\expandafter\let\csname endequation*\endcsname\relax

\usepackage{amsmath,amssymb,amsfonts}
\usepackage{algorithmic}
\usepackage{graphicx, caption}
\usepackage{textcomp}
\usepackage{lineno}

\usepackage{graphicx}
\usepackage{dcolumn}
\usepackage{bm}

\usepackage[utf8]{inputenc}
\usepackage[T1]{fontenc}
\usepackage{mathptmx} 

\begin{document}

\title{Unifying Consciousness and Time to Enhance Artificial Intelligence}
\author{Mahendra Samarawickrama}
\email[]{samarawickrama@gmail.com}
\affiliation{Centre for Consciousness Studies, Australia}
\date{\today}


\begin{abstract}
Consciousness is a sequential process of awareness which can focus on one piece of information at a time. This process of awareness experiences causation which underpins the notion of time while it interplays with matter and energy, forming reality. The study of Consciousness, time and reality is complex and evolving fast in many fields, including metaphysics and fundamental physics. Reality composes patterns in human Consciousness in response to the regularities in nature. These regularities could be physical (e.g., astronomical, environmental), biological, chemical, mental, social, etc. The patterns that emerged in Consciousness were correlated to the environment, life and social behaviours followed by constructed frameworks, systems and structures. The complex constructs evolved as cultures, customs, norms and values, which created a diverse society. In the evolution of responsible AI, it is important to be attuned to the evolved cultural, ethical and moral values through Consciousness. This requires the advocated design of self-learning AI aware of time perception and human ethics. 
\end{abstract}

\maketitle

\section{Introduction}
\begin{figure*}[ht!]
  \centering
  \includegraphics[width=\textwidth]{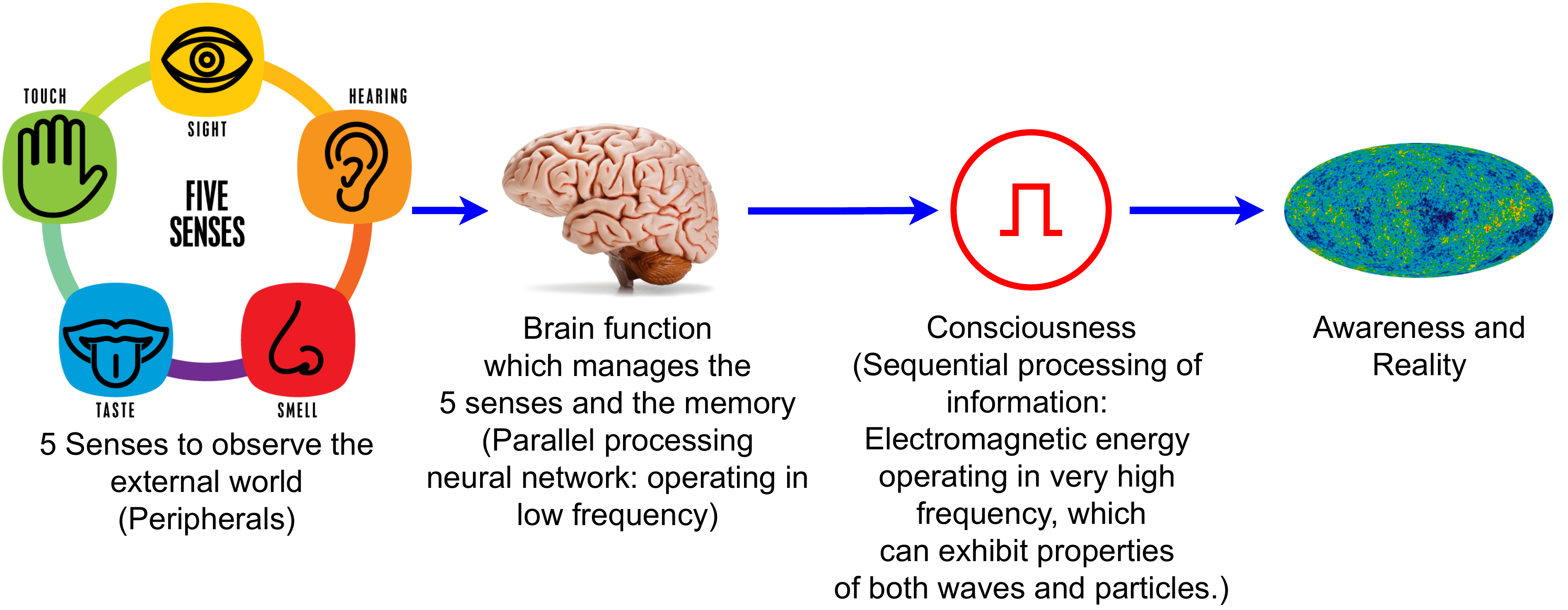}
  \caption{The interplay of five sensors, brain and consciousness. The brain processes sensory information in parallel. However, the awareness of causation (i.e., consciousness) is a sequential process focusing on a single piece of information at a time. This sequential process of awareness in consciousness operates fast and consistently, which underpins our perception of reality.}
  \label{brain_consciousness_model}
\end{figure*}

The notion of time is an integral part of consciousness \cite{10.1093/nc/niab015}. The consciousness experiences the causation or changes in reality/environment and perceives the time. Therefore, in our previous publication \cite{Samarawickrama2022}, we assumed that consciousness is a sequential process which is aware of a single piece of information at a time. Even though the brain processes sensory data of five sensors (i.e., Sight, Sound, Smell, Taste, and Touch) in parallel in the neural network, the awareness of causation is a sequential process following cause and effect. See the illustration of this idea in Figure~\ref{brain_consciousness_model},
 
The assumption of sequential awareness in consciousness enables mapping the perception of time into consciousness. Based on the theory of relativity \cite{Einstein:1905:EBK}, the perception of time is relative to the frame of reference. Einstein assumed that the speed of light is constant in all frames of reference, and the time is derived based on that fundamental assumption. In our paper, we defined the shortest time to be aware of reality as a consciousness cycle. Then based on relativity, this consciousness cycle is also subjected to dilation, like relativistic time \begin{equation}
\label{consciousness_dilation}
T_v = \frac{T_0}{\sqrt{1-\frac{v^2}{c^2}}}\,,
\end{equation}
where, $T_v$ is the dilated period of the consciousness cycle related to the rest period of the consciousness cycle $T_0$. Note that the $\sqrt{1-\frac{v^2}{c^2}}$ is the \textit{Lorentz factor}, where $v$ is the relative velocity between inertial reference frames, and $c$ is the speed of light in a vacuum. Then, we mathematically modelled \cite{Samarawickrama2022} how consciousness would interplay with matter and energy, forming reality, which can be adapted to understand limitations and opportunities in AI consciousness. This paper extends our discussion towards the time perception of artificial intelligence systems (AIS). 

\section{The notion of time in perception and reality}

Humans, like any other life forms, experience time through causation. Patterns are composed in the human consciousness in response to the regularities in nature \cite{Blackburn1990}. Since the beginning of human civilisation, humans have learnt and evolved complex concepts and constructs by incorporating time emerged through patterns in the consciousness. The earth’s rotation around itself determines the day, and orbiting around the sun determines the year. The Moon takes about one month to orbit the earth. The tilt of the earth’s spin axis with respect to its orbital plane causes the weather seasons. These environmental patterns cause many biological patterns and lifestyle patterns in human life. To predict and organise these patterns effectively, humans introduce standard time with clocks, calendars and various other frameworks. These artificial frameworks enable us to model time and objectively measure subjective experiences. 

Physics has been evolved by observation of nature with various frameworks of time. In this way, time became an essential construct and dimension of our understanding of reality. For example, Newtonian physics \cite{newton1687} evolved assuming that time is absolute and flows consistently from past to present and into the future. That enables the development of mathematical models for explaining patterns in reality with time. However, later observations, such as the perihelion motion of Mercury, allow humans to understand time as a relativistic measure rather than an absolute. The modern understanding of the universe is based on the theory of relativity \cite{Einstein:1915:ARG, Einstein:1915:EPM}, which is completely articulated by space-time principles. Based on relativity, John Wheeler \cite{Wheeler:1961:GPM} stated, “Space tells matter how to move. Matter tells space how to curve”. Relativity enables us to accurately understand and predict the behaviours of black holes, stars, and planets. Further, relativity enables humans to develop technologies like the atomic clock \cite{Ramsey_2005} and Global Positioning System (GPS) \cite{Maddison2009} that are useful in everyday life.

The behaviour of particles is completely different to larger objects like planets, stars, etc. This led to the evolution of Quantum physics \cite{10.1093/ptep/ptaa104} as opposed to relativity. Quantum physics exhibits amazing accuracy in predicted results in particle physics. However, it greatly disturbs the notion of time modelled in relativity. For example, in the collapse of the wave function in quantum entanglement, Einstein described that as a spooky action at a distance \cite{EPR1935}. As per relativity, information cannot transfer faster than the speed of light. As per the recent discoveries in quantum entanglement, information can be transferred instantly, faster than the speed of light, making our reality non-local \cite{10.1080/19420889.2020.1822583}. The non-local reality contradicts relativity, which is now applied in quantum teleportation at the subatomic level. On the other hand, at the quantum level, the reality is uncertain, as described by Heisenberg’s uncertainty principle \cite{Heisenberg:1927:AIQ}. As per the uncertainty principle, it is impossible to precisely measure or be aware of the position and speed of a particle in a given time. This brings the limitation of human awareness and perception of time. Therefore, many believe now that consciousness is fundamental and that time and causation are derived from consciousness \cite{Hoffman2014_Time}.

\section{The implication of principles of time for AIS}

The inability to consolidate quantum physics and the theory of relativity makes our understanding of reality incomplete. Moreover, the new discoveries proving the idea of non-local reality shake the status quo of fundamental physics \cite{10.3389/fphy.2020.00360}. Therefore, it is still impossible to supervise AI to experience the notion of time to understand reality precisely. On the other hand, human understanding of reality is also about 5\%, whereas most of the universe consists of dark matter and dark energy, which humans do not understand \cite{OKS2021101632}. Under these conditions, AI might be used to explore reality and time in a way we have never imagined. Perhaps incorporating AI to understand reality and causation might help humans to become fully aware of reality by overcoming inherent biases from evolution, culture and nature. 

Typical Reinforcement Learning (RL) technique can be adapted to automate the learning of AI. The RL process can be mathematically formulated using Markov Decision Process (MDP) \cite{vanOtterlo2012}. That is a sequential learning process by trial and error. In this process, the learning agent (i.e., AI) sequentially interacts with the environment with an intelligent decision (i.e. action) followed by receiving a reward or a penalty based on the policy imposed. There will be no influence on the AI agent’s action, but convey the value of its action through feedback with reward or penalty. This way, the AI agent will self-learn about the environment over time. The RL process is illustrated in Figure~\ref{RL_MDP}:
\begin{figure*}[ht!]
  \centering
  \includegraphics[width=\textwidth]{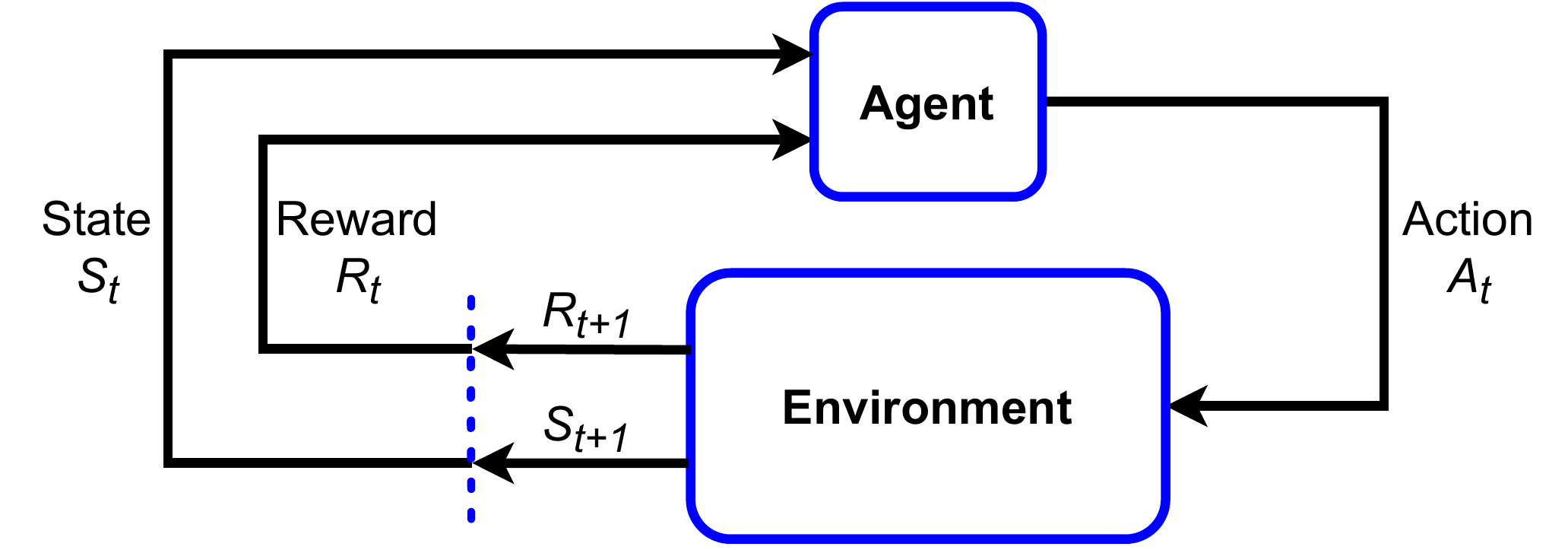}
  \caption{Components of the Markov Decision Process (MDP) and its function in the agent-environment interaction. The sequential step of time is represented by $t$.}
  \label{RL_MDP}
\end{figure*}

\section{The implication of human beliefs, values and cultures for the perception of time in AIS}

Human beliefs, customs, culture and values are tightly linked with various dynamics and interpretations of the time and periodicities based on the movement of the earth, Moon and other terrestrial bodies. From the beginning, humans identified that time affects life and nature differently. Therefore, in the Greece era, early Western culture, there were at least three gods representing different time forms: \textit{Chronos}, \textit{Aion}, and \textit{Kairos} \cite{10.2307/27902109}. \textit{Chronos} represented the linear time flowing from past to present into the future. This is the time that humans feel when life passes. In contrast, \textit{Aion} represented the cyclical nature of time experienced from natural events such as weather patterns, rebirths, etc. The third god \textit{Kairos} represented the opportunist time, which reflects the appropriate time to achieve a task. In this way, time, environment and beliefs were tightly linked with life and governed society and values.

On the other hand, in Eastern culture, the horoscope is one good example of a planetary and constellation framework underpinning Astrology as a foundation of certain belief systems \cite{Campion2015}. These beliefs assume that Astrology is associated with time and causality, which can predict the future and guide humans. 

The human observation of the night sky led to perceiving time from various cyclical patterns going far back in time. For example, the Aboriginal Australians \cite{Hamacher2008} observed the night sky and mapped them to the environment and life stages that evolved various customs, arts and even religions. Not only by interacting planets and stars but the tilt of the earth’s spin axis also significantly led to diversifying human cultures based on seasons, particularly when moving away from the equator. 

The notion of time and associated beliefs, customs, and values are important to consider when training AIS \cite{10.1080/1097198X.2020.1794131}. That will help promote human cultural values, ethics, and diversity, equity and inclusion (DEI). AI development may need to pay attention to and integrate the time attributes that emerged from nature, values and cultures. Humans may include them in the policies for rewarding self-learning AI algorithms (e.g., in MDP).

\section{The implication of biological time on AIS}

The biological cycles play a fundamental role in human behaviours and the perception of time—for example, mood cycles, circadian rhythms, and the menstrual cycle. Without understanding these biological time-keeping processes, AI cannot seamlessly integrate with human society when creating values in health, culture, art, etc. These insights are essential to realising emotional intelligence, empathy and awareness in AI. Literature shows the effective use of Cyclic Hidden Markov Models (CyH-MMs) for detecting and modelling cycles in a multidimensional heterogeneous biological time series data collection \cite{10.1145/3178876.3186052}. It is important to attribute the relevant features of biological processes when training AIS, which raises more awareness about humans. 

Recent discoveries in quantum physics argue that our reality is non-local, where awareness can happen instantly, faster than the speed of light. Physicists and neurologists think brain neurons might be aware of the quantum world through the orchestrated collapse of microtubules in the neurons in the brain \cite{Penrose2014, 10.1080/17588928.2020.1839037}. If this hypothesis is true, then there are possibilities that human awareness can be linked with non-local realities to expand our consciousness across the universe instantly. From this perspective, future AI might need to be evolved with the capabilities of biological neurons, which interplay with the quantum realities. The recent development of neurotech realising brain-computer interface (BCI) along with emerging quantum computers might enable such capabilities in the near future \cite{Saha2021}.

\section{Conclusion}

Consciousness and perception of time and causation are key to awareness and understanding reality. The notion of time emerged from causation, a perception relative to the observer as per the relativity principles. In relativity, it’s not time but the light-speed constant in all frames of reference. In contrast, in quantum entanglement, the reality is non-local, and information can be transferred instantly faster than light. While the principles of time contradict the foundation of physics, time also influenced the formation of diverse customs, values and cultures based on patterns that emerged from nature, particularly around the regularities in the earth’s movement, environment, astronomy and biology. Therefore,  understanding time and related artefacts (i.e., cultures, beliefs, values, customs, physics, health, etc.) are very important to realise deep awareness of reality. From the AIS perspective, it will enhance the understanding of AI in human health, cultures, customs, values and various other diversities. Bringing this awareness to AI will be a challenging and complex yet rewarding milestone in the evolution of ethical and responsible AI.

\bibliographystyle{ieeetr}
\bibliography{references}

\end{document}